\documentclass{iau} 
\usepackage{graphicx}

\newcommand{\Msun}{{\rm ~M}_{\odot}}

\newcommand{\Rsun}{{\rm ~R}_{\odot}}
\newcommand{\Zsun}{{\rm ~Z}_{\odot}}
\newcommand{\gpy}{{\rm ~Gpc}^{-3} {\rm ~yr}^{-1}}

\title[~~Local merger rates of double neutron stars] 
{Local merger rates of double neutron stars}

\author[Martyna Chruslinska ]   
{Martyna Chruslinska$^1$
}

 \affiliation{$^{1}$Department of Astrophysics/IMAPP, Radboud University,
 P.O. Box 9010, NL-6500 GL Nijmegen, The Netherlands \\ email: {\tt m.chruslinska@astro.ru.nl} \\[\affilskip]
}

\pubyear{2018}
\volume{346}  
\jname{High-mass X-ray binaries: illuminating the passage from
massive binaries to merging compact objects}
\editors{A.C. Editor, B.D. Editor \& C.E. Editor, eds.}
\begin{document}

\maketitle

\begin{abstract}
The first detection of gravitational waves from a merging double neutron star (DNS) binary 
 implies a much higher rate of DNS coalescences in the local Universe
 than typically estimated on theoretical grounds.
The recent study by Chruslinska et al.(2018) shows that
 apart from being particularly sensitive to the common envelope treatment, DNS merger rates appear rather robust against
 variations of several factors probed in their study 
 (e.g. conservativeness of the mass transfer, angular momentum loss, and natal kicks), unless extreme assumptions are made.
 Confrontation with the improving observational limits may allow to rule out some of the extreme models.
To correctly compare model predictions with observational limits
 one has to account for the other factors that affect the rates.
 One of those factors relates to the assumed history of star formation 
 and chemical evolution of the Universe and its impact on the final results needs to be better constrained. 

\end{abstract}

\firstsection 
\section{Introduction}

\begin{figure}[ht]
\begin{center}
 \includegraphics[width=3.8in]{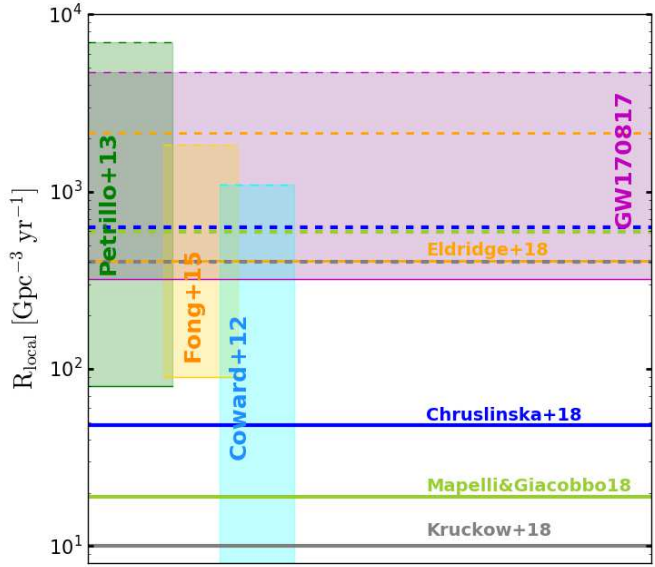} 
 \caption{Shaded areas show observational limits on double neutron star (DNS) local merger rate density ($R_{loc}$)
  implied by the detection of GW170817 in gravitational waves (purple; \cite[Abbott et al. 2017a]{GW170817}) 
  and based on short gamma ray burst observations
  from Coward et al. 2012 (cyan), Petrillo et al. 2013 (green) and Fong et al. 2015 (yellow).
  Thick lines mark the recent population synthesis results obtained by different groups 
  (blue - Chruslinska et al. 2018a, light green - Mapelli \& Giacobbo 2018,
  gray - Kruckow et al. 2018, orange - Eldridge et al. 2018).
  The solid lines show results for models indicated by the authors as fiducial, while the dashed ones
  show the highest DNS $R_{loc}$ found within each study.}
   \label{fig: rates}
\end{center}
\end{figure}

When two stars composing a binary system complete their evolution they leave behind two compact remnants orbiting each other
- so called double compact object (DCO). If the system evolves in isolation (dynamical interactions can be neglected), 
its subsequent orbital evolution is determined by the angular momentum loss due to gravitational wave emission.
If the orbital separation is small enough, two stellar remnants merge within the Hubble time (Peters 1964).
The current network of ground based gravitational wave detectors can observe the final stages of mergers
of DCOs composed of neutron stars (NS) and black holes (BH) happening in the local Universe 
(\cite[Abbott et al. 2016]{Abbott16_limits}).
With the increasing number of detections and observing time those observations will allow to put progressively tighter 
constraints on the frequency of mergers of different types of DCOs occurring in the probed volume 
(local merger rate density $R_{loc}$).
The first detection of gravitational waves from a merging double neutron star (DNS) (\cite[Abbott et al. 2017a]{GW170817}) 
confirmed that those events are responsible for the production of at least some of the short gamma ray bursts (sGRB) and kilonovae 
(\cite[Abbott et al. 2017b]{GW170817_multimess}). 
In principle, electromagnetic observations of these phenomena can be also used to set limits on $R_{loc}$ for DNS.
However, the weakly constrained sGRB jet opening angle and luminosity function (e.g. \cite[Petrillo et al. 2013]{Petrillo13})
and small number of observations in case of kilonovae 
lead to substantial uncertainties and differences in the limits estimated by different groups (see Fig. \ref{fig: rates}).
Moreover, a (unknown) fraction of those events may originate from NS-BH/BH-NS mergers (e.g. \cite[Berger 2014]{Berger14}).
Thus, gravitational waves provide the most direct measure of cosmological DNS merger rates.
\\
One can estimate $R_{loc}$ on theoretical grounds following the evolution of stars either in isolation 
(e.g. \cite[Tutukov \& Yungelson 1993]{TutukovYungelson93}; \cite[Portegies Zwart \& Yungelson 1998]{ZwartYungelson98};
\cite[Belczynski et al. 2002]{Belczynski02})
or in dense environments 
(e.g. \cite[Potregies Zwart et al. 2004]{Zwart04}; \cite[Rodriguez et al. 2016]{Rodriguez16}; \cite[Askar et al. 2017]{Askar17}), 
where dynamical interactions are important, and studying what fraction of them produces DCOs
close enough to merge within the Hubble time.
In this contribution we focus on the isolated evolution.
The rates in this channel are usually calculated using the population synthesis method.
In this approach one follows the evolution of many systems from the zero age main sequence (ZAMS) phase,
when their parameters (mass of the primary star, mass ratio, separation, eccentricity) are randomly drawn from
observation-based distributions (\cite[Sana et al. 2012]{Sana12}; \cite[Moe \& Di Stefano 2017]{MoeDiStefano17}).
During the subsequent evolution the binary parameters change due to
wind mass loss (e.g. \cite[Vink et al. 2001]{Vink01}), interaction between the stars via mass transfer or tides, 
possible mass loss from the system during the formation of each of the compact objects
and natal kick velocity gained by NS or BH due to asymmetries involved in the process of their formation 
(e.g. \cite[Gunn \& Ostriker 1970]{GunnOstriker70}; \cite[Hobbs et al. 2005]{Hobbs05};
\cite[Fryer \& Kushenko 2006]{FryerKushenko06}; \cite[Janka 2017]{Janka17}).
Only a small fraction of the simulated binaries reach the interesting DCO stage
(many systems merge during the unstable mass transfer phases
or get disrupted during the formation of one of the compact objects)
and have parameters allowing for their merger within the Hubble time (e.g. \cite[Chruslinska et al. 2018a]{Chruslinska18a}).
This fraction (and hence the merger rate) depends on the choice of a particular set 
of assumptions (parameters that define a model) used to describe the weakly constrained phases of binary evolution, 
e.g. conservativeness of the mass transfer, distribution describing the magnitude of NS and BH natal kicks
(e.g. \cite[Mennekens \& Vanbeveren 2014]{MennekensVanbeveren14};
\cite[Chruslinska et al. 2018a]{Chruslinska18a}; \cite[Kruckow et al. 2018]{Kruckow18}; 
\cite[Mapelli \& Giacobbo 2018]{MapelliGiacobbo18}; \cite[Eldridge et al. 2018]{Eldridge18}; 
\cite[Barrett et al 2018]{Barrett18}).
\\
Confrontation of the calculated rates with observational limits is one of the possible ways to constrain those models,
hopefully allowing to rule out some of the extreme models and pointing us towards the correct understanding of the binary evolution.
However, to be meaningful this comparison must take into account also other factors, besides those related directly to binary evolution, 
that affect the estimated $R_{loc}$. 
Those include the choice of distributions describing initial parameters of binary stars 
and formation and evolution of progenitor stars in the chemically evolving Universe.
While the uncertainty caused by the former has been already studied by \cite[de Mink \& Belczynski (2015)]{deMinkBelczynski15}
and \cite[Klencki et al. (2018)]{Klencki18},
the impact of the latter still lacks discussion in the literature.

\section{Binary evolution and DNS merger rates}\label{sec: evolution}

\begin{figure}[ht]
\begin{center}
 \includegraphics[width=5.5in]{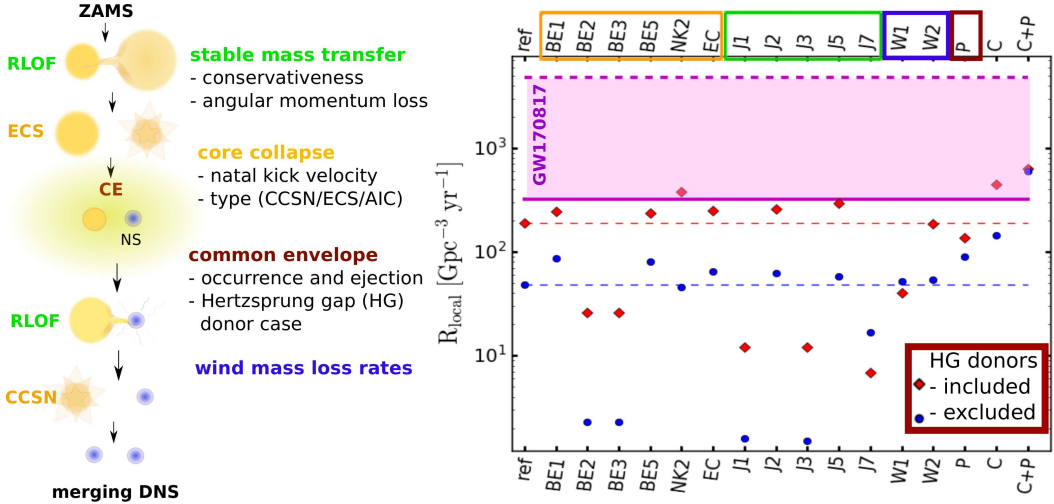} 
 \caption{ \underline{Left:} schematic picture of the evolution leading to formation of a merging double neutron star (DNS).
  This path involves several stages of stable mass transfer (RLOF), two core-collapse events leading to formation of both NS
  (ECS-electron capture supernova, CCSN - core collapse supernova) and common envelope evolution (CE).
  The major evolution-related factors that affect $R_{loc}$ calculations and need to be parametrized in population synthesis 
  studies at each of these stages are listed in the middle.
  \newline
  \underline{Right:} DNS $R_{loc}$ for different population synthesis models.
  Colors on the top axis indicate which evolutionary phase was parametrized differently with respect to the reference ($ref$) model 
  (orange - core collapse, green - mass transfer, brown - common envelope). 
  Models $C$ and $C+P$ involve combinations of several modifications favoring the formation of merging DNS.
  Each model comes in two variations depending on whether evolution through CE with HG type donors was allowed (red diamonds) or
  excluded, assuming to lead to merger (blue dots).
  The shaded purple region indicates observational limits implied by detection of GW170817.
  Modified figure from Chruslinska et al. (2018a). See the original paper for details.
  }
   \label{fig: schema}
\end{center}
\end{figure}

The impact of binary evolution-related assumptions on merger rates of double neutron star systems has been 
studied by many groups over the years, with the use of different population synthesis codes
(e.g. \cite[Tutukov \& Yungelson 1993]{TutukovYungelson93}; \cite[Belczynski et al. 2002]{Belczynski02}; 
\cite[Mennekens \& Vanbeveren 2014]{MennekensVanbeveren14};
\cite[Chruslinska et al. 2018a]{Chruslinska18a}; \cite[Kruckow et al. 2018]{Kruckow18}; 
\cite[Mapelli \& Giacobbo 2018]{MapelliGiacobbo18}; \cite[Eldridge et al. 2018]{Eldridge18}; 
\cite[Barrett et al 2018]{Barrett18}).
The reported $R_{loc}$ span more than an order of magnitude and fall between a few 10 $\gpy$ to a few 100 $\gpy$ 
when the fiducial models are considered. In this contribution we focus only on the most recent studies.
In Figure \ref{fig: rates} we show the most recent population synthesis results
together with observational limits on DNS $R_{loc}$ resulting from detection of gravitational waves from 
merging DNS (GW170817; \cite[Abbott et al. 2017a]{GW170817}) and short gamma ray bursts observations.
The short GRB limits come from \cite[Coward et al. (2012)]{Coward12}, \cite[Petrillo et al. (2013)]{Petrillo13} and \cite[Fong et al. (2015)]{Fong15}.
In case of \cite[Petrillo et al. (2013)]{Petrillo13} the upper limit assumes the jet beaming angle of 10$^\circ$ and the lower limit assumes
weakly collimated emission of 60$^{\circ}$ (see Figure 3 therein). 
The very conservative lower limit from \cite[Coward et al. (2012)]{Coward12} assumes isotropic sGRB jet emission.
As shown in Fig. \ref{fig: rates}, most of the population synthesis results are either below or fall on the
lower side of the current observational limits, even despite the broad range of reported values.
The wide range of results coming from those studies reflects our limited knowledge of the details of the evolution of massive stars in binaries.
The typical evolutionary path for the formation of a merging DNS together with the major unknown factors in the evolution is summarized 
in Figure \ref{fig: schema}.
Below we provide a short overview of the most important factors that affect the evolution towards a merging DNS. For more details
see sec. 2 in \cite[Chruslinska et al. (2018a)]{Chruslinska18a} and references therein.
\\
The final DNS must end up on the very close orbit of a few solar radii to belong to the merging population (\cite[Peters et al. 1964]{Peters64}).
Thus, its progenitor stars form on the orbit that ensures interaction via mass transfer phase(s)
when one of the components over-fills its Roche lobe
(RLOF; unless a forming NS receives a very favorably oriented natal kick velocity that efficiently decreases the orbital separation).
The amount of material lost from the system during the mass transfer (conservativeness) 
and angular momentum carried away with the escaping mass are
one of the unknowns in binary evolution
(e.g. \cite[de Mink et al. 2007]{deMinkPolsHilditch07}).
\\
Furthermore, formation of DNS involves two core-collapse events in which NS form.
These can be either iron-core collapse supernovae (CCSN), electron-capture supernovae 
(ECS; when the partially degenerate ONeMg core of mass$\sim 1.37 \Msun$ collapses due to electron captures on Mg and Ne;
e.g. \cite[Miyaji et al. 1980]{Miyaji80})
or accretion induced collapse (AIC; e.g. \cite[Nomoto \& Kondo 1991]{NomotoKondo91}) of a massive accreting white dwarf to a NS.
The exact conditions for the occurrence of different core collapse events, 
especially the boundary between ECS and CCSN are currently not known (e.g. \cite[Jones et al. 2016]{Jones16}).
Binary interaction is believed to broaden the range of masses of progenitor stars that can undergo ECS 
(e.g. \cite[Podsiadlowski et al. 2004]{Podsiadlowski04}; \cite[van den Heuvel 2007]{vdHeuvel07}), 
which in case of single stars is very narrow 
(and hence their evolution through ECS is unlikely).
Neutron stars forming due to electron-capture triggered collapse (either ECS or AIC) are believed to gain relatively small natal kick velocities 
($\lesssim$50km/s; e.g. \cite[van den Heuvel 2007]{vdHeuvel07}).
Taking into account that the typical birth velocities of young single pulsars are of the order of a few 100 km/s
(e.g. \cite[Hobbs et al. 2005]{Hobbs05}),
this formation path significantly increases the chance that a binary will remain bound during the NS formation. 
It has been suggested that the magnitude of natal kick may depend on the amount of mass ejected during the supernova explosion
(with smaller mass of ejecta leading to smaller birth velocity; 
e.g. \cite[Beniamini \& Piran 2016]{BeniaminiPiran16}; \cite[Bray \& Eldridge 2016]{BrayEldridge16}; \cite[Janka et al. 2017]{Janka17};
\cite[Bray \& Eldridge 2018]{BrayEldridge18}).
Thus, NS progenitors that loose a large fraction of their envelope during mass transfer phases
may potentially form with smaller natal kick velocities.
Such severe envelope stripping can occur when the companion star is already a compact object,
leading to ultra-stripped supernova phenomenon
(e.g. \cite[Tauris et al. 2013]{Tauris13}; \cite[Tauris et al. 2015]{Tauris15}).
However, the mechanism responsible for production of natal kicks is not well understood.
If the role of neutrino asymmetries is dominant (e.g. \cite[e.g. Fryer \& Kushenko 2006]{FryerKushenko06}), the suggested
ejecta mass - natal kick correlation may turn out erroneous. 
The distribution of natal kicks is one of the key ingredients of population synthesis studies
focusing on DNS. Different assumptions can significantly affect the size of the final population and hence the calculated merger rates.
\\
Finally, the classical formation channel of merging DNS requires at least one stage of 
unstable mass transfer (so-called common envelope; CE e.g. \cite[Ivanova et al. 2013]{Ivanova13}).
The radii of massive stars typically reach 100 - 1000 $\Rsun$, 
which is much larger than the required separation for the merging systems
(a few solar radii for DNS binary). 
To explain the formation of such systems without dynamical interactions, one needs the mechanism capable of
decreasing the orbital separation by even a few orders of magnitude.
Common envelope is believed to provide such a mechanism.
CE forms when the mass transfer rate is too high for the accretor to accrete all of the transferred material, 
giving rise to a short-lived phase during which both stars are immersed in a shared envelope. 
This causes a binary inspiral due to increased friction and, if the envelope is not ejected beforehand (e.g. at the expense of the orbital energy),
may lead to its coalescence before the formation of a DNS.
This evolutionary phase is particularly weakly constrained both by theory and observations (e.g. \cite[Ivanova et al. 2013]{Ivanova13}). 
Neither the conditions for the onset of CE, nor for its ejection are presently known. 
Both are likely dependent on the structure and type of stars involved
and require detailed modeling. The particularly controversial case arises when 
the donor star does not have a well defined core-envelope boundary,
as for the stars during the Hertzsprung gap phase (referred to as HG type donors; e.g. \cite[Ivanova \& Taam 2004]{IvanovaTaam04}). 
Such CE was believed to lead to binary merger, however recent
studies by \cite[Pavlovskii \& Ivanova (2015)]{PavlovskiiIvanova15} and \cite[Pavlovskii et al. (2017)]{Pavlovskii17}
show that in some cases where based on earlier studies one would expect the common envelope initiated by a
HG type donor to ensue, the mass transfer may in fact be stable. 

\subsection{Recent population synthesis results}

\cite[Chruslinska et al. (2018a)]{Chruslinska18a} revisited the topic of DNS merger rates in light of
recent observational and theoretical progress in the study of the evolution of double compact objects.
They conclude that apart from being particularly sensitive to the common envelope treatment and 
extreme assumptions about natal kicks, 
DNS $R_{loc}$ appear rather robust against variations of several of the key factors probed in their study. 
Within 21 models calculated with the \textsc{StarTrack} population synthesis code 
(\cite[Belczynski et al. 2002]{Belczynski02}; \cite[Belczynski et al. 2008]{Belczynski08}) they identify only two variations
leading to significant (a factor of $\sim$10) increase in $R_{loc}$ of NS-NS binaries with respect to the reference model. 
These are the models that combine several factors supporting the formation of double neutron star binaries,
requiring simultaneous changes in the treatment of common envelope evolution,
angular momentum loss, natal kicks and electron capture supernovae.
The summary of $R_{loc}$ for models considered in this study is shown in the right panel of Figure \ref{fig: schema}.
Only three of the presented models fall above the LIGO/Virgo lower 90\% confidence limit on DNS $R_{loc}$ 
implied by the detection of GW170817 (\cite[Abbott et al. 2017a]{GW170817}).
At the same time the local BH-BH merger rate densities calculated for those three models are found to exceed
the corresponding gravitational waves limits (\cite[Abbott et al. 2016]{Abbott16_limits}). 
A possible solution to this discrepancy, as suggested by the authors,
might be different CE evolution in case of the most massive stars (BH-progenitors) than in case of NS-progenitors
or higher BH natal kicks than assumed in the simulations, 
both factors largely affecting the BH-BH merger rate calculations and currently weakly constrained.
We return to this question in section \ref{sec: other}.
\\
A number of other groups published their results on DCO local merger rates during the last year.
The highest NS-NS $R_{loc}$ of $\sim$600 $\gpy$ quoted by \cite[Mapelli \& Giacobbo (2018)]{MapelliGiacobbo18}  
(dashed green line in Fig. \ref{fig: rates}) 
was obtained assuming 5 times higher efficiency of CE ejection than in their reference model and additionally 
requiring that all NS form with low natal kicks 
(drawn from Maxwellian distribution with velocity dispersion $\sigma$=15 km/s). Note that these velocities are 
generally lower than in any of the models
presented in Chruslinska et al. (2018). The highest $R_{loc}$ found by the latter is also around 600 $\gpy$.
\cite[Kruckow et al. (2018)]{Kruckow18} find 'optimistic' DNS $R_{loc}$ of $\sim$400 $\gpy$ also
requiring low natal kicks at NS formation. $R_{loc}$ of $\sim$400 $\gpy$ was also found by
\cite[Eldridge et al. (2018)]{Eldridge18} for their fiducial model.
The highest NS-NS $R_{loc}$ found by \cite[Eldridge et al. (2018)]{Eldridge18} reaches up to $\sim 2100 \gpy$
and was obtained using their updated prescription for NS natal kicks (\cite[Bray \& Eldridge 2018]{BrayEldridge18}).
In this prescription natal kick velocity scales with the amount of mass ejected during the supernova and mass of the stellar
remnant (see discussion in \cite[Janka 2017]{Janka17}), as opposed to the commonly used approach in which those velocities are
randomly drawn from observationally motivated distribution (e.g. \cite[Hobbs et al. 2005]{Hobbs05}).
They find that for the earlier version of this natal kick model (\cite[Bray \& Eldridge 2016]{BrayEldridge16})
the estimated NSNS local rates reasonably agree with the results obtained for the corresponding model from \cite[Chruslinska et al. (2018a)]{Chruslinska18a}
that employed the prescription from \cite[Bray \& Eldridge (2016)]{BrayEldridge16}.
\\
All of the above values are consistent with the 90\% confidence limits on DNS $R_{loc}$ 
implied by the detection of GW170817 (1540$^{+3200}_{-1220}\gpy$).
However, except for the highest value reported by \cite[Eldridge et al. (2018)]{Eldridge18} 
for the model using the \cite[Bray \& Eldridge (2018)]{BrayEldridge18} natal kick prescription, 
they fall decidedly on the lower side of this estimate.

\section{Other factors affecting calculation of the local merger rates}\label{sec: other}

\subsection{Distributions of initial parameters of binaries}

Before drawing conclusions from the comparison with observations, 
one has take into account other factors, besides the evolution-related assumptions, that affect the population synthesis results.
One of those factors relates to the initial conditions of the simulations that have the form of distributions 
describing parameters of binaries at their formation. 
\cite[de Mink \& Belczynski (2015)]{deMinkBelczynski15} studied how the properties of merging DCOs
change when they use initial conditions based on study of massive spectroscopic binaries (\cite[Sana et al. 2012]{Sana12}) instead of 
the previously used distributions of initial parameters. They report a slight increase (a factor of $\lesssim$2) in the DNS merger rates, 
mostly due to higher binary fraction (fraction of stars assumed to form in binaries). They argue that the uncertainty in the high-mass
slope of the initial mass function (IMF) may affect the results by a factor of $\sim$4, however, 
recent study by \cite[Klencki et al. (2018)]{Klencki18} shows that this effect is in fact less significant 
if the assumed star formation rate (SFR) and IMF are varied consistently.
\cite[Klencki et al. (2018)]{Klencki18} implement the empirical inter-correlated distributions of initial binary parameters
reported by \cite[Moe \& Di Stefano (2017)]{MoeDiStefano17} based on their analysis of over twenty massive star surveys.
The variations in the merger rates due to this change stays within a factor of $\sim$2 when compared with the 
simulations using initial distributions from \cite[Sana et al. (2012)]{Sana12}. 
This transition generally decreases the merger rate estimates for DNS. 
This effect is the strongest at high metallicities.
\subsection{Chemical evolution of the Universe}

\begin{figure}[ht]
\begin{center}
 \includegraphics[width=3.8in]{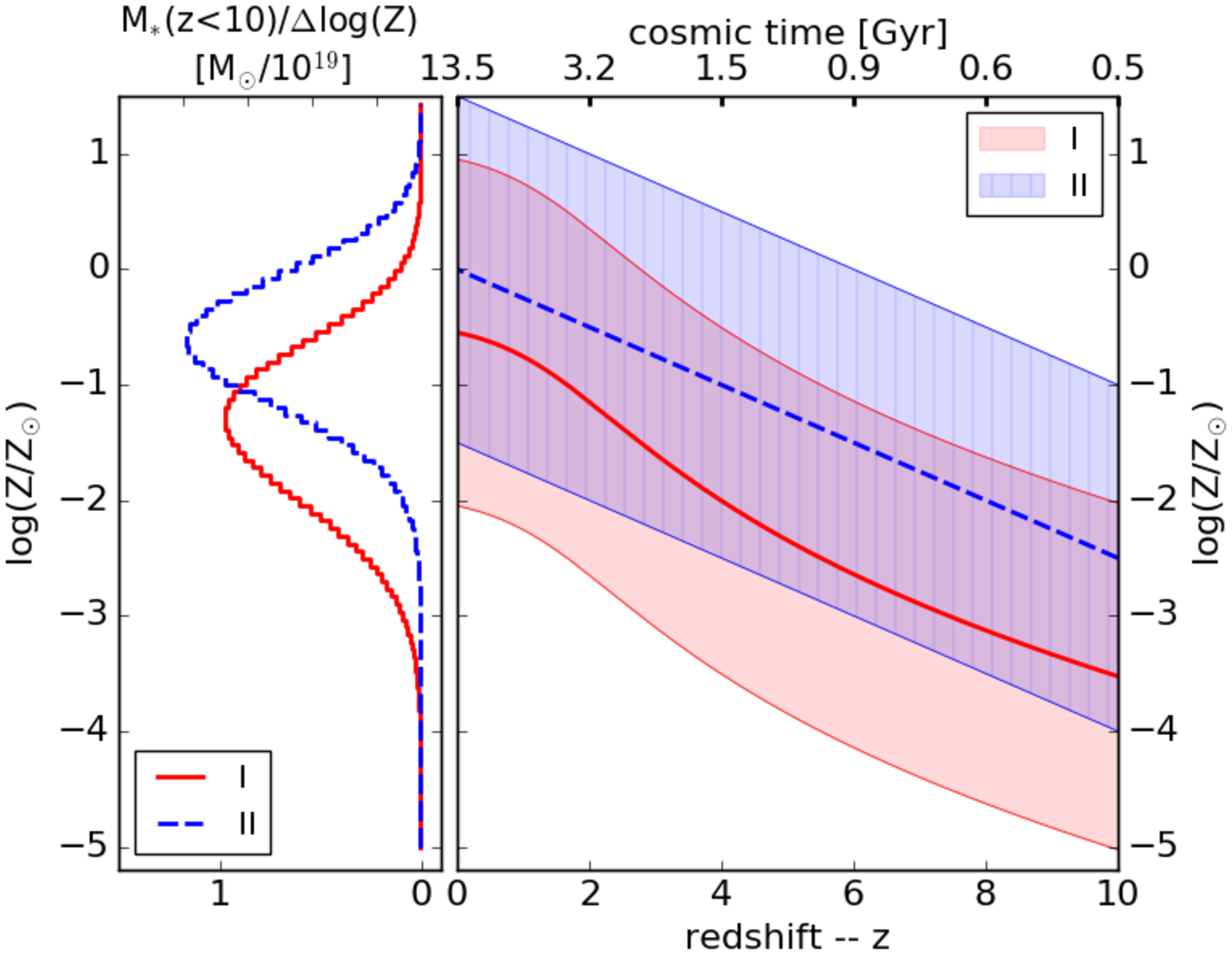} 
 \caption{ 
 \underline{Right:} comparison of redshift evolution of the mean metallicity as assumed by \cite[Belczynski et al. (2016)]{Belczynski16N}
 ($I$) and \cite[Dominik et al. (2013)]{Dominik13} ($II$). The shaded areas indicate 3$\sigma$ spread limits around the mean.
 \newline
 \underline{Left:} 
  distribution of mass formed in stars since redshift z=10 at different metallicities for both versions of the metallicity evolution.
  The fraction of mass formed at low metallicity ($<0.1 \Zsun$) since redshift of 10 amounts to 71\% in case $I$ and 27\% in case $II$.
  Modified figure from \cite[Chruslinska et al. (2018b)]{Chruslinska18b}.
  }
   \label{fig: Zave}
\end{center}
\end{figure}

Metallicity itself is one of the crucial factors determining the evolution of stars. 
It affects for instance wind mass loss rates and stellar radii, also impacting the evolution of stars in binaries 
and the outcome of their evolution (e.g. \cite[Maeder 1992]{Meader92}; \cite[Hurley et al. 2000]{Hurley00}; 
\cite[Vink et al. 2001]{Vink01}; \cite[Belczynski et al. 2010]{Belczynski10}).
In particular, the number of close double compact binaries of a certain type created per unit of mass formed in stars 
(DCO formation efficiency $\rm \chi_{DCO;i}$) 
is known to vary significantly depending on the composition of progenitor stars 
(e.g. \cite[Dominik et al. 2012]{Dominik12}; \cite[Eldridge \& Stanway 2016]{EldridgeStanway16}; 
\cite[Stevenson et al. 2017]{Stevenson17}; \cite[Klencki et al. 2018]{Klencki18}; \cite[Giacobbo et al. 2018]{Giacobbo18}). 
\\
A double compact object that merges in the local Universe (at time t$_{mr}$) 
forms at some earlier time t$_{\rm form}$=t$_{mr}$-t$_{del}$-t$_{DCO}$,
where t$_{DCO}$ is time needed to complete binary evolution up to the formation of a DCO 
and t$_{del}$ time needed to decrease its separation
due to gravitational wave emission to the point where two stellar remnants merge. 
Its progenitor binary forms with metallicity that is typical for
its neighborhood at time t$_{form}$ 
\footnote{The metallicity of the star-forming material evolves over time
due to interplay between metal enrichment by evolving stars, 
feedback from supernovae and active galactic nuclei and possible
inflows of metal-poor material.}.
Since DCOs form with different parameters, they have a range of t$_{del}$. As a consequence,
the local merger rate density is a summed contribution of merging DCOs that formed at
different cosmic times and with different metallicities.
Hence, $R_{loc}$ depends on both the amount of star formation happening throughout the cosmic time 
(which sets the number of progenitor binaries formed at a given time) and on its distribution across different metallicities
($\rm SFR(t,Z)$; because the fraction of progenitor binaries that end up as merging DCOs of certain type depends on metallicity).
Furthermore, the dependence of $\rm \chi_{DCO;i}$ on metallicity is different for different types of DCOs 
(e.g. Fig. 6 in \cite[Klencki et al. (2018)]{Klencki18}; more (recent) SFR at low metallicities generally
favors formation of merging BH-BH while discouraging the formation of merging DNS) and the exact form of this dependence 
is generally sensitive to the population synthesis assumptions (\cite[Chruslinska et al. 2018b]{Chruslinska18b}). 
As a result, different assumptions about $\rm SFR(t,Z)$ affect the ratio of $R_{loc}$ calculated for different types of DCOs
and the uncertainty associated with this choice can vary between the models. 
\\ \newline
Different approaches have been taken to determine $\rm SFR(t,Z)$ used to calculate merger rate densities.
 One can extract this information from cosmological simulations 
 (e.g. \cite[Mapelli et al. 2017]{Mapelli17}; \cite[Schneider et al. 2017]{Schneider17}), 
 or use the available observations and/or complement observational results with theoretical inferences
 ( e.g. \cite[Dominik et al. 2013]{Dominik13}; \cite[Belczynski et al. 2016]{Belczynski16N}; 
  \cite[Eldridge et al. 2018]{Eldridge18}; Chruslinska et al. in prep.).
 All methods have their weaknesses.
 Observations are prone to biases and provide complete information only in limited ranges of redshifts and luminosities of the objects of interest.
 On the other hand, cosmological simulations do not fully account for all of the observational relations 
 (e.g. mass - metallicity relation) and are resolution-limited. 
  In any case, the use of incorrect SFR(t,Z) affects the resulting cosmological merger rates and may lead to erroneous conclusions.
  However, the importance of this choice was not evaluated in previous studies.
  \\ \newline
  Following the approach described in \cite[Belczynski et al. (2016)]{Belczynski16N},  \cite[Chruslinska et al. (2018a)]{Chruslinska18a}
  used  the cosmic star formation rate density from \cite[Madau \& Dickinson (2014)]{MadauDickinson14} 
  and mean metallicity of the Universe as a function of redshift from the  chemical  evolution  model proposed by these authors.
  This metallicity was increased by 0.5 dex and a Gaussian spread of $\sigma$=0.5 dex was added to the mean to construct
  a distribution describing contributions from different metallicities to SFR at different times.   
  However, observations of star-forming galaxies in the local Universe suggest that
  the mean metallicity at which star formation takes place is likely higher than assumed by \cite[Belczynski et al. (2016)]{Belczynski16N}.  
  Massive galaxies (with stellar masses $\gtrsim10^{9}\Msun$) showing SFR even two orders of magnitude higher than in their dwarf counterparts,
  dominate the star formation budget in the Universe 
  (e.g. \cite[Brinchmann et al. 2004]{Brinchmann04}; \cite[Lara-L{\'o}pez et al. 2013]{LaraLopez13}; \cite[Boogard et al. 2018]{Boogaard18}).
  The star forming gas found in those galaxies has relatively high metal content, which is close to the solar value $\Zsun$,
  even if uncertainty in the absolute metallicity calibration is taken into account (e.g. \cite[Kewley \& Ellison 2008]{KewleyEllison08}).
  Thus, the amount of low-metallicity ($\lesssim 0.1 \Zsun$) star formation assumed in 
  \cite[Chruslinska et al. (2018a)]{Chruslinska18a} may be overestimated (see also discussion in Klencki et al. 2018 and Chruslinska et al. 2018b).
  A more detailed study of the SFR(t,Z) resulting from currently available observations of star-forming galaxies
  and the associated uncertainties is underway.\\
  To demonstrate the potential effect of the assumed SFR(t,Z) on their merger rate density calculations,
  we contrast the assumptions made by \cite[Belczynski et al. (2016)]{Belczynski16N} and \cite[Dominik et al. (2013)]{Dominik13}.
  In Figure \ref{fig: Zave} we show the mean metallicity evolution with redshift used in both cases.
  In the method proposed by \cite[Dominik et al. (2013)]{Dominik13} 
  the average metallicity of the star formation in the local Universe is $\sim \Zsun$ (in contrast to $\sim0.3 \Zsun$ in Belczynski et al. 2016)
  and the amount of mass formed in stars at metallicity $\lesssim 0.1\Zsun$ since redshift of 10 is $\sim$2.6 times lower 
  when compared with the method proposed by \cite[Belczynski et al. (2016)]{Belczynski16N}. 
  For the model with DCO formation efficiency dependence on metallicity as shown in Fig. 6 from \cite[Klencki et al. (2018)]{Klencki18} 
  (identical to the reference model from Chruslinska et al. 2018a) 
  this difference translates to a factor of $\sim 1.5$ increase of $R_{loc}$ for DNS and a factor of $\sim$4 decrease in $R_{loc}$ for BH-BH.
  As shown by \cite[Chruslinska et al. (2018b)]{Chruslinska18b} this difference is also sufficient to resolve the reported discrepancy between 
  the gravitational waves limits on BH-BH $R_{loc}$ and $R_{loc}$ calculated for those systems in model $CA$ from 
  \cite[Chruslinska et al. (2018a)]{Chruslinska18a} which led to DNS $R_{loc}$ that satisfy current limits implied by the detection of GW170817.
  In this model several moderate modifications were introduced to their reference model, all found to favor the formation
  of merging DNS: 
   \cite{BrayEldridge16} prescription for the natal kicks, 
  reduced angular momentum loss during the mass transfer and wider limits
  on the helium core mass for the progenitors of stars undergoing electron-capture supernovae.
  Furthermore, the evolution through a common envelope phase initiated by a Hertzsprung gap donor was allowed.
  In this model the discrepancy can be resolved by the use of SFR(t,Z) closer to that expected based on observations of local
  star-forming galaxies,
  without the need for different evolution-related assumption in the BH- and NS-progenitors mass regime. 
  \\
  This highlights the importance of the choice of a particular way to distribute the cosmic star formation rate across metallicities and time 
  and the need to better understand the uncertainties associated with that choice. 
  Without tighter constraints on SFR(t,Z) one has to deal with another layer of degeneracy in the calculated merger rates, 
  which makes drawing any strong conclusions from studies that aim to use cosmological rates as constraints impossible.
  
\section{Conclusions}

The first detection of gravitational waves from merging double neutron star binary implied
local merger rate density ($R_{loc}$) of this type of systems that is much higher than most current
theoretical estimates. While certain population synthesis models can produce rates consistent
with the 90\% confidence limits reported after this event, they fall on the low side of this estimate
unless specific assumptions about natal kicks at the formation of NS are used.
If the future observations support the rate close to the current most likely value reported by LIGO/Virgo,
our understanding of the evolution of massive stars in binaries may need revision.\\
At the same time, confrontation of the theoretical results with observational limits
requires knowledge of the uncertainties induced by other factors, besides those related
to assumptions about the binary evolution that are accounted for in the population synthesis parameter studies.
\\
One of those factors, necessary to calculate volumetric rates of any stellar evolution-related phenomena,
relates to the assumed fraction of star formation happening at different metallicities throughout the cosmic time SFR(t,Z).
Because of different dependence of the efficiency of formation of different types of merging double compact objects on metallicity
the choice of SFR(t,Z) strongly affects the estimated ratio of merger rates of different types of binaries.
Different assumptions about SFR(t,Z) were made in the literature, however the importance of those choices has not been discussed.
The SFR(t,Z) used by \cite[Chruslinska et al. (2018a)]{Chruslinska18a} likely overestimates the amount of low metallicity star formation.
While a detailed study of the SFR(t,Z) resulting from observational properties of star-forming galaxies is underway,
to demonstrate the possible impact of this assumption on the final results, we estimate the change in the calculated NS-NS and BH-BH
rates caused by the use of SFR(t,Z) that assumes higher metallicity of the star formation in the local Universe.
This change results in a factor of 1.5 increase in NS-NS $R_{loc}$ and a factor of 4 decrease in BH-BH $R_{loc}$ for a model
with DCO formation efficiency - metallicity relation as in their reference model and is sufficient to affect 
some of the conclusions drawn from their study (\cite[Chruslinska et al. 2018b]{Chruslinska18b}).
\\
This highlights the importance of the choice of a particular way to distribute the cosmic star formation rate across metallicities
and time and the need to better understand the uncertainties associated with that choice.


\begin{thebibliography}{}
\bibitem[Abbott et al. (2016)]{Abbott16_limits}
{Abbott, B.~P., Abbott, R., Abbott, T.~D., Abernathy, M.~R., Acernese, F., Ackley, K.,
  Adams, C., Adams, T., Addesso, P., Adhikari, R.~X. et al.} 2016,
\textit{ApJ}, 832, L21

\bibitem[Abbott et al. (2017a)]{GW170817}
{Abbott, B.~P., Abbott, R., Abbott, T.~D., Acernese, F., Ackley, K., Adams, C., Adams, T.,
 Addesso, P., Adhikari, R.~X., Adya, V.~B. et al.} 2017a,
 \textit{Phys. Rev. Lett.}, 119, 161101

\bibitem[Abbott et al. (2017b)]{GW170817_multimess} 
{Abbott, B.~P., Abbott, R., Abbott, T.~D., Acernese, F., Ackley, K., Adams, C., Adams, T.,
 Addesso, P., Adhikari, R.~X., Adya, V.~B. et al.} 2017b,
\textit{Ap. Lett.}, 848, L12 
 
\bibitem[Askar et al. (2017)]{Askar17}
{Askar, A., Szkudlarek, M., Gondek-Rosinska, D., Giersz, M. \& Bulik, T.} 2017,
\textit{MNRAS}, 464, L36
 
 \bibitem[Barrett et al. (2018)]{Barrett18}
 {Barrett, J. W., Gaebel, S. M., Neijssel, C. J., Vigna-Gómez, A.,
 Stevenson, S., Berry, C. P. L., Farr, W. M. \& Mandel, I.} 2018,
\textit{MNRAS}, 477, 4685
 
\bibitem[Belczynski et al. (2002a)]{Belczynski02a}
{ Belczynski K., Kalogera V. \& Bulik T.} 2002,
\textit{ApJ} 572, 407
 
\bibitem[Belczynski et al. (2008)]{Belczynski08} 
{Belczynski K., Kalogera V., Rasio F. A., Taam R. E., Zezas A.,
 Bulik T., Maccarone T. J. \& Ivanova N.} 2008, 
 \textit{ApJS}, 174, 223
 
\bibitem[Belczynski et al. (2010)]{Belczynski10}
{Belczynski, K., Bulik, T., Fryer, C.~L., Ruiter, A., Valsecchi, F., Vink, J.~S. \& Hurley, J.~R.} 2010,
\textit{ApJ}, 714, 1217

\bibitem[Belczynski et al. (2016)]{Belczynski16N}
{Belczynski, K., Holz, D.~E., Bulik, T. \& O'Shaughnessy, R.} 2016,
\textit{Nature}, 534, 512

\bibitem[Beniamini \& Piran (2016)]{BeniaminiPiran16}
{Beniamini P. \& Piran T.} 2016, 
\textit{MNRAS}, 456, 4089

\bibitem[Berger (2014)]{Berger14}
{Berger, E.} 2014,
\textit{ARA\&A}, 52, 43

\bibitem[Bray \& Eldridge (2016)]{BrayEldridge16}
{Bray J. C. \& Eldridge J. J.} 2016, 
\textit{MNRAS}, 461, 3747

\bibitem[Bray \& Eldridge (2018)]{BrayEldridge18}
{Bray J. C. \& Eldridge J. J.} 2018, 
\textit{MNRAS}, 480, 5657

\bibitem[Brinchmann et al. (2004)]{Brinchmann04}
{Brinchmann, J., Charlot, S., White, S.~D.~M., Tremonti, C., Kauffmann, G., Heckman, T. \& Brinkmann, J.} 2004,
\textit{MNRAS}, 351, 1151

\bibitem[Boogaard et al. (2018)]{Boogaard18}
{Boogaard, L.~A., Brinchmann, J., Bouch{\'e}, N., Paalvast, M., 
 Bacon, R., Bouwens, R.~J., et al.} 2018,
 \textit{ArXiv e-prints}, arXiv:1808.04900

\bibitem[Chruslinska et al. (2018a)]{Chruslinska18a}
{Chruslinska, M., Belczynski, K., Klencki, J., \& Benacquista, M.} 2018a,
\textit{MNRAS}, 474, 2937

\bibitem[Chruslinska et al. (2018b)]{Chruslinska18b}
{Chruslinska, M., Nelemans, G. \& Belczynski, K.} 2018b,
\textit{ArXiv e-prints}, arXiv:1811.03565

\bibitem[Coward et al. (2012)]{Coward12}
{Coward D. M., et al.} 2012, 
\textit{MNRAS}, 425, 2668

\bibitem[de Mink et al. (2007)]{deMinkPolsHilditch07}
{de Mink S. E., Pols O. R. \& Hilditch R. W.} 2007, 
\textit{A\&A}, 467, 1181

\bibitem[de Mink \& Belczynski (2015)]{deMinkBelczynski15}
{de Mink S. E. \& Belczynski K.} 2015, 
\textit{ApJ}, 814, 58

\bibitem[Dominik et al. (2012)]{Dominik12}
{ Dominik M., Belczynski K., Fryer C., Holz D. E., Berti E., Bulik
T., Mandel I. \& O’Shaughnessy R.} 2012, 
\textit{ApJ}, 759, 52

\bibitem[Dominik et al. (2013)]{Dominik13}
{Dominik M., Belczynski K., Fryer C., Holz D. E., Berti E., Bulik
T., Mandel I. \& O’Shaughnessy R.} 2013, 
\textit{ApJ}, 779, 72

\bibitem[Eldridge \& Stanway (2016)]{EldridgeStanway16}
{Eldridge J. J. \& Stanway E. R.} 2016, 
\textit{MNRAS}, 462, 3302

\bibitem[Eldridge et al. (2018)]{Eldridge18}
{Eldridge J. J., Stanway E. R. \& Tang P. N.} 2018,
\textit{ArXiv e-prints}, arXiv:1807.07659

\bibitem[Fong et al. (2015)]{Fong15}
{Fong W., Berger E., Margutti R. \& Zauderer B. A.} 2015, 
\textit{ApJ}, 815, 102

\bibitem[Fryer \& Kushenko (2006)]{FryerKushenko06}
{Fryer C. L. \&  Kusenko A.} 2006, 
\textit{ApJS}, 163, 335

\bibitem[Giacobbo et al. (2018)]{Giacobbo18}
{Giacobbo N., Mapelli M. \& Spera M.} 2018, 
\textit{MNRAS}, 474, 2959

\bibitem[Gunn \& Ostriker (1970)]{GunnOstriker70}
{Gunn J. E. \& Ostriker J. P.} 1970, 
\textit{ApJ}, 160, 979

\bibitem[Hobbs et al. (2005)]{Hobbs05}
{Hobbs G., Lorimer D. R., Lyne A. G. \& Kramer M.} 2005, 
\textit{MNRAS}, 360, 974

\bibitem[Hurley et al. 2000]{Hurley00}
{Hurley, J.~R., Pols, O.~R. \& {Tout}, C.~A.} 2000,
\textit{MNRAS},315, 543

\bibitem[Ivanova \& Taam (2004)]{IvanovaTaam04}
{Ivanova N.\& Taam R. E.} 2004, 
\textit{ApJ}, 601, 1058

\bibitem[Ivanova et al. (2013)]{Ivanova13}
{Ivanova N. et al.} 2013, 
\textit{A\&AR}, 21, 59

\bibitem[Janka (2017)]{Jakna17}
{Janka H.-T.} 2017, 
\textit{ApJ}, 837, 84

\bibitem[Jones et al. (2016)]{Jones16}
{Jones S., Ropke F. K., Pakmor R., Seitenzahl I. R., Ohlmann
S. T. \& Edelmann P.~V.~F.} 2016, 
\textit{A\&A}, 593, A72

\bibitem[Kewley \& Ellison (2008)]{KewleyEllison08}
{Kewley, L.~J. \& Ellison, S.~L.} 2008,
\textit{ApJ}, 681, 1183

\bibitem[Klencki et al. (2018)]{Klencki18}
{Klencki, J., Moe, M., Gladysz, W., Chruslinska, M., Holz, D.~E. \& Belczynski, K.} 2018,
\textit{ArXiv e-prints}, arXiv:1808.07889

\bibitem[Kruckow et al. (2018)]{Kruckow18}
{Kruckow, M.~U., Tauris, T.~M., Langer, N., Kramer, M. \& Izzard, R.~G.} 2018,
\textit{MNRAS}, 481, 1908

\bibitem[Lara-L{\'o}pez et al. (2013)]{LaraLopez13}
{Lara-L{\'o}pez, M. A., Hopkins, A. M., L{\'o}pez-S{\'a}nchez, A. R., et al.} 2013,
\textit{MNRAS}, 434, 451

\bibitem[Madau \& Dickinson (2014)]{MadauDickinson14}
{Madau, P. \& Dickinson, M.} 2014,
\textit{ARA\&A}, 52, 415

\bibitem[Mapelli et al. (2017)]{Mapelli17}
{Mapelli, M., Giacobbo, N., Ripamonti, E., \& Spera, M.} 2017,
\textit{MNRAS}, 472, 2422

\bibitem[Mennekens \& Vanbeveren (2014)]{MennekensVanbeveren14}
{Mennekens N. \& Vanbeveren D.} 2014,
\textit{A\&A}, 564, A134

\bibitem[Meader (1992)]{Meader92}
{Maeder, A.} 1992,
\textit{A\&A}, 264, 105

\bibitem[Miyaji et al. (1980)]{Miyaji80}
{Miyaji S., Nomoto K., Yokoi K. \& Sugimoto D.} 1980, 
\textit{PASJ}, 32, 303

\bibitem[Moe \& Di Stefano (2017)]{MoeDiStefano17}
{Moe, M. \& Di Stefano, R.} 2017, 
\textit{ApJS}, 230, 15

\bibitem[Nomoto \& Kondo (1991)]{NomotoKondo91}
{Nomoto K. \& Kondo Y.} 1991,
\textit{ApJ}, 367, L19

\bibitem[Pavlovskii \& Ivanova (2015)]{PavlovskiiIvanova15}
{Pavlovskii K. \& Ivanova N.} 2015, 
\textit{MNRAS}, 449, 4415

\bibitem[Pavlovskii et al. (2017)]{Pavlovskii17}
{Pavlovskii K., Ivanova N., Belczynski K. \& Van K. X.} 2017, 
\textit{MNRAS}, 465, 2092

\bibitem[Peters (1964)]{Peters64}
{Peters, P.~C.} 1964,
\textit{Phys. Rev.}, 136, 1224

\bibitem[Petrillo et al. (2013)]{Petrillo13}
{Petrillo C. E., Dietz A. \& Cavaglia M.} 2013, 
\textit{ApJ}, 767, 140

\bibitem[Podsiadlowski et al. (2004)]{Podsiadlowski04}
{Podsiadlowski P., Langer N., Poelarends A. J. T., Rappaport S.,
Heger A. \& Pfahl E.} 2004, 
\textit{ApJ}, 612, 1044

\bibitem[Portegies Zwart \& Yungelson (1998)]{ZwartYungelson98}
{Portegies Zwart, S.~F. \& Yungelson, L.~R.} 1998,
\textit{A\&A}, 332, 173

\bibitem[Portegies Zwart et al. (2004)]{Zwart04}
{Portegies Zwart, S. F., Baumgardt, H., Hut, P., Makino, J. \& McMillan, S. L. W.} 2004, 
\textit{Nature}, 428, 724

\bibitem[Rodriguez et al. (2016)]{Rodriguez16}
{Rodriguez, C. L., Haster, C.-J., Chatterjee, S., Kalogera, V. \& Rasio, F. A.} 2016, 
\textit{ApJ}, 824, L8

\bibitem[Sana et al. (2012)]{Sana12}
{Sana H., et al.} 2012, 
\textit{Science}, 337, 444

\bibitem[Schneider et al. (2017)]{Schneider17}
{Schneider, R., Graziani, L., Marassi, S., Spera, M., Mapelli, M., Alparone, M., \& Bennassuti, M.~d.} 2017,
\textit{MNRAS}, 471, L105

\bibitem[Stevenson et al. (2017)]{Stevenson17}
{Stevenson S., Vigna-Gómez A., Mandel I., Barrett J. W., Neijssel C. J.,
Perkins D. \& de Mink S. E.} 2017, 
\textit{Nature Communications}, 8, 14906

\bibitem[Tauris et al. (2013)]{Tauris13}
{Tauris T. M., Langer N., Moriya T. J., Podsiadlowski P., Yoon
S.-C., Blinnikov S. I.} 2013, 
\textit{ApJ}, 778, L23

\bibitem[Tauris et al. (2015)]{Tauris15}
{Tauris T. M., Langer N. \& Podsiadlowski P.} 2015, 
\textit{MNRAS}, 451, 2123

\bibitem[Tutukov \& Yungelson (1993)]{TutukovYungelson93}
{Tutukov, A. V. \& Yungelson, L. R.} 1993, 
\textit{MNRAS}, 260, 675

\bibitem[van den Heuvel (2007)]{vdHeuvel07}
{van den Heuvel E. P. J.} 2007,
in: di Salvo T., Israel G. L., Piersant L., Burderi L., Matt G., Tornambe A. \& Menna M. T. (eds),
\textit{The Multicolored Landscape of Compact Objects and Their Explosive Origins}
AIP Conf. Ser. Vol. 924, (Am. Inst. Phys., New York), p.\ 598

\bibitem[Vink et al. (2001)]{Vink01}
{Vink, J.~S., de Koter, A. \& Lamers, H.~J.~G.~L.~M.}, 2001
\textit{A\&A}, 369, 574


\end{thebibliography}
\end{document}